\documentclass{iopjournal}

\usepackage{bm,graphicx,textcomp,amssymb,amsmath,dcolumn,color}
\usepackage{hyperref}

\newcommand{\ket}[1]{\left|#1\right>}
\newcommand{\bra}[1]{\left<#1\right|}
\newcommand{\braket}[2]{\left<#1|#2\right>}
\newcommand{\nn}{\nonumber\\}

\newcommand{\f}[1]{\mbox{\boldmath$#1$}}

\newcommand{\bea}{\begin{eqnarray}}
\newcommand{\ea}{\end{eqnarray}}
\newcommand{\eea}{\end{eqnarray}}
\newcommand{\ord}{\,{\cal O}}

\newcommand{\expval}[1]{\langle #1 \rangle}
\newcommand{\ii}{{\rm i}}
\newcommand{\abs}[1]{\left| #1 \right|}
\newcommand{\trace}[1]{{\rm Tr}\left\{#1\right\}}

\newcommand{\new}[1]{{#1}}

\begin{document}
\articletype{Original Research Article}

\title{Quantum Zeno effect versus adiabatic quantum computing and quantum annealing}

\author{Naser Ahmadiniaz$^1$, Dennis Kraft$^1$, Gernot Schaller$^1$, and Ralf Sch\"utzhold$^{1,2}$}

\affil{$^1$ Helmholtz-Zentrum Dresden-Rossendorf, 
Bautzner Landstra{\ss}e 400, 01328 Dresden, Germany}

\affil{$^2$ Institut f\"ur Theoretische Physik, 
Technische Universit\"at Dresden, 01062 Dresden, Germany}

\email{g.schaller@hzdr.de}


\date{\today}

\begin{abstract}
For the adiabatic version of Grover's quantum search algorithm as proposed 
by Roland and Cerf, we study the impact of decoherence caused by a rather 
general coupling to some environment. 
For quite generic conditions, we find that the quantum Zeno effect poses 
strong limitations on the performance (quantum speed-up) since the environment
effectively measures the state of the system permanently and thereby inhibits 
or slows down quantum transitions. 
Generalizing our results, we find that \new{similar} 
restrictions should apply universally to adiabatic quantum algorithms 
and quantum annealing schemes which are based on 
\new{analogous} 
isolated Landau-Zener type transitions at avoided level 
crossings (similar to first-order phase transitions). 
As a possible resort, more gradual changes of the quantum state 
(as in second-order phase transitions) 
or suitable error-correcting schemes such as the spin-echo method 
may alleviate this problem. 
\end{abstract}


\section{Introduction}

\subsection{Quantum Zeno Effect}\label{Quantum Zeno Effect}

The Greek philosopher Zeno (or Zenon) formulated several apparent paradoxes 
illuminating the tension between the two concepts of being in motion versus 
being at a certain position at a given time. 
As a famous example, imagine a race between the fast Achilles and a slow 
tortoise where both start at the same time but the tortoise gets a head start.
Then, before being able to overtake the tortoise, Achilles must first reach 
the place where the tortoise started -- but, during that time, the tortoise 
already moved away and covered an additional distance. 
Thus, Achilles needs some more time to cover this distance too, 
which allows the tortoise to move even further, and so on. 
This leads to a never ending sequence such that one might expect that 
Achilles should never be able to overtake the tortoise.

Intriguingly, this apparent paradox resurges in quantum theory~\cite{misra1977a,itano1990a,hackenbroich1998a,fischer2001a,gurvitz2003a,streed2006a,bernu2008a,slichter2016a,blumenthal2022a,ahmadiniaz2022a}
where frequent measurements of the state of a system 
(corresponding to the concept of position) may slow down or even 
inhibit the quantum evolution in the form of a transition to another state 
(corresponding to the concept of motion). 
As a simple example, let us consider the unitary evolution of a spin-$1/2$ 
system $\ket{\psi(t)}$ which rotates 
\new{
$\ket{\psi(t)}=\cos(\omega t)\ket{\uparrow}+\sin(\omega t)\ket{\downarrow}$
} 
from the initial state $\ket{\uparrow}$ to the final state $\ket{\downarrow}$.
Measuring the state of the system 
(in the $\ket{\uparrow},\ket{\downarrow}$ basis) 
after a short time \new{$\Delta t\ll1/\omega$} 
projects it to one of the basis states with the probabilities 
$P_\uparrow=\cos^2(\new{\omega}\Delta t)$
or 
$P_\downarrow=\sin^2(\new{\omega}\Delta t)\approx(\new{\omega}\Delta t)^2$.
Repeating this measurement $N$ times with $N\new{\omega}\Delta t=\pi/2$ 
then yields the total transition probability 
$1-\cos^{2N}(\new{\omega} \Delta t)$ which scales as $N(\new{\omega}\Delta t)^2\ll1$,
i.e., the transition rate is strongly reduced from $\ord(\new{\omega})$ to 
$\ord(\new{\omega}^2\Delta t)$. 
As one way to understand this reduction, we have to sum small 
probabilities (in the case of measurements) instead of small 
amplitudes (for unitary evolution). 

Besides the projective measurements discussed above, weak measurements~\cite{aharonov2014a,kastner2017a}
can have a similar effect~\cite{breuer2007,wiseman2010} -- provided that the system dynamics is slow 
enough. 
In order to illustrate this point, let us consider the standard 
Lindblad master equation ($\hbar=1$) 
\begin{align}
\label{EQ:master}
\frac{d\hat\rho}{dt}=-i\left[\hat H,\hat\rho\right]+
\hat L\hat\rho\hat L^\dagger-
\frac12\left\{\hat L^\dagger\hat L,\hat\rho\right\} 
\,,
\end{align}
where the system Hamiltonian $\hat H$ generates the unitary time 
evolution of the density matrix $\hat\rho$ while the Lindblad (jump) 
operator $\hat L$ describes the impact of the environment, which may 
cause a non-unitary time evolution. 

Choosing a spin-$1/2$ system with 
$\hat H=\new{\omega}\hat\sigma_y$,
the unitary system dynamics itself 
(i.e., for $\hat L=0$) would correspond to the rotation between the 
states $\ket{\uparrow}$ and $\ket{\downarrow}$ as described above. 
Now let us add the Lindblad operator $\hat L=\sqrt{\Gamma}\,\hat\sigma_z$ 
which generates decoherence with the damping rate $\Gamma$ and leads to a 
decay of the coherences between the states $\ket{\uparrow}$ and $\ket{\downarrow}$, 
i.e., the off-diagonal matrix elements of $\hat\rho$ 
in the $\ket{\uparrow},\ket{\downarrow}$ basis (phase damping). 
Then the evolution equation for the expectation value of the spin 
operator $\langle\f{\hat\sigma}\rangle$ reads 
\begin{align}\label{EQ:masterspec}
\frac{d}{dt}\langle\f{\hat\sigma}\rangle=
-2\f{\new{\omega}}\times\langle\f{\hat\sigma}\rangle
+2\f{L}\times\left(\f{L}\times\langle\f{\hat\sigma}\rangle\right) 
\,,
\end{align}
with $\f{\new{\omega}}=\new{\omega}\f{e}_y$
and $\f{L}=\sqrt{\Gamma}\,\f{e}_z$. 
For $\new{\omega}\gg\Gamma$, we approximately find \new{a} rotation with a 
frequency $\new{\omega}$ which slowly decays with a damping rate of $\ord(\Gamma)$.
For $\new{\omega}\ll\Gamma$, on the other hand, the rotation disappears and 
we find a slow decay to the fully mixed state $\hat\rho=\f{1}/2$ (with $\f{1}$ denoting the unit matrix)
on a long time scale of $\ord(\Gamma/\new{\omega}^2)$. 
As an intuitive picture, the environment measures the state of the system 
with an effective measurement rate $\Gamma\sim1/\Delta t$.
Thus, the slow decay to $\hat\rho=\f{1}/2$ with the decay rate 
$\ord(\new{\omega}^2/\Gamma)$ corresponds to the example above with the 
reduced transition rate $\ord(\new{\omega}^2\Delta t)$. 
Note that analogous results can be obtained for Lindblad operators in the 
form of projectors $\hat L=\sqrt{\Gamma}\ket{\uparrow}\bra{\uparrow}$ 
or $\hat L=\sqrt{\Gamma}\ket{\downarrow}\bra{\downarrow}$
or a combination of both 
$\hat L_1=\sqrt{\Gamma_1}\ket{\uparrow}\bra{\uparrow}$ 
and $\hat L_2=\sqrt{\Gamma_2}\ket{\downarrow}\bra{\downarrow}$, 
see Appendix~\ref{SEC:lindblad_slow}.
Furthermore, in App.~\ref{SEC:general_slow} we also provide an example of Zeno freezing in the strong-coupling regime that does not require a Lindblad treatment.

\new{Summarizing the above results, the quantum Zeno effect can be induced 
actively~\cite{kitano1997a,novelli2015a,jamadagni2021a,burgarth2022a} by performing strong (projective) or weak measurements on the 
system or passively by coupling the system to an environment which 
effectively monitors the state of the system. 
In all cases, an important point is the competition between the 
measurement or decoherence rate $\Gamma$ and the frequency $\new{\omega}$ 
of the internal system dynamics~\cite{facchi2002a,cresser2006a,maniscalco2006a,bosco_de_magalhaes2011a,wu2017a}.}

\subsection{Adiabatic Quantum Algorithms}
\label{Adiabatic Quantum Algorithms}

In contrast to the usual scheme of sequential quantum algorithms~\cite{nielsen2000} where the 
desired final state $\ket{\psi_{\rm out}}$
is obtained by applying a sequence of quantum gates onto 
some initial state $\ket{\psi_{\rm in}}$, 
adiabatic quantum algorithms~\cite{vandam2001a,farhi2001a,albash2018a} 
encode the solution to the 
problem of interest 
into the ground state of a suitably chosen 
Hamiltonian $\hat H_{\rm out}$. 
In order to reach this final ground state $\ket{\psi_{\rm out}}$,
the idea is to start with an initial Hamiltonian $\hat H_{\rm in}$ 
whose ground state $\ket{\psi_{\rm in}}$ is known and can be prepared 
efficiently and then to slowly deform the initial Hamiltonian 
$\hat H_{\rm in}$ into the final $\hat H_{\rm out}$.
A simple example is a linear interpolation scheme 
\begin{align}
\label{linear-interpolation}
\hat H(t)=f(t)\hat H_{\rm in}+\left[1-f(t)\right]\hat H_{\rm out}
\,,
\end{align} 
with some function $f(t)$ satisfying $f(t=0)=1$ and $f(t=T)=0$
where $T$ is the run-time of the algorithm. 

If the change from $\hat H_{\rm in}$ to $\hat H_{\rm out}$ is slow 
enough, the adiabatic theorem ensures that we indeed end up in the 
desired final ground state $\ket{\psi_{\rm out}}$. 
More precisely, an important condition for adiabatic evolution is that 
the matrix element of the rate of change $\partial_t\hat H $ between 
two different instantaneous eigen-vectors 
$\hat H(t)\ket{\psi_n(t)}=E_n(t)\ket{\psi_n(t)}$ with instantaneous 
eigen-energies $E_n(t)$ is small~\cite{sarandy2004a,jansen2007a}
\begin{align}
\label{EQ:adiabatic} 
\abs{\bra{\psi_n}\partial_t\hat H\ket{\psi_m}}\ll(E_n-E_m)^2 
\,.
\end{align}
Assuming adiabatic evolution and starting in an eigen-vector 
$\ket{\psi(t=0)}=\ket{\psi_n(t=0)}$ such as the ground state $n=0$, 
the time evolved quantum state $\ket{\psi(t)}$ stays close to the 
instantaneous eigen-vector $\ket{\psi_n(t)}$ up to phase factors 
containing the dynamical phase 
$\dot\varphi_n^{\rm dyn}(t)=E_n(t)$ and the geometrical phase 
$\dot\varphi_n^{\rm geo}(t)=\ii \bra{\psi_n(t)}\partial_t\ket{\psi_n(t)}$. 

\subsection{Adiabatic Quantum Search Algorithm}
\label{Adiabatic Quantum Search Algorithm}

As a simple example for such an adiabatic quantum algorithm, let us consider 
for a single solution the adiabatic version of Grover's quantum search algorithm~\cite{grover1997a} as proposed by 
Roland and Cerf~\cite{roland2002a}.
In the Grover search, the task is to find the solution state $\ket{w}$ 
as one of the $N$ computational basis states (such as $\ket{w}=\ket{01\dots10}$) 
within an unsorted database. 
Since we obviously do not know the solution $\ket{w}$ 
beforehand, 
we choose the superposition state 
$\ket{s}=\sum_{x=0}^{N-1}\ket{x}/\sqrt{N}$
as the initial state. 
In an effective spin representation with $n$ spins and $N=2^n$ basis states 
and identifying $\ket{\uparrow}=\ket{1}$ 
and $\ket{\downarrow}=\ket{0}$, this state 
$\ket{s}=[(\ket{0}+\ket{1})/\sqrt{2}]^{\otimes n}$ 
has all spins aligned in $x$-direction 
$\ket{s}=\ket{\rightarrow}\ket{\rightarrow}\dots\ket{\rightarrow}$. 

A simple realization of the linear interpolation 
scheme~\eqref{linear-interpolation} is then 
\begin{align}\label{EQ:adiabatic_search_algorithm}
\hat H(t)=-\Omega\left[
f(t)\ket{s}\bra{s}+\left[1-f(t)\right]\ket{w}\bra{w}
\right] 
\,.
\end{align} 
In this simple example, the non-trivial quantum evolution is restricted 
to the subspace spanned by the vectors $\ket{w}$ and $\ket{s}$ or 
$\ket{w}$ and $\ket{w_\perp}=\ket{s}-\ket{w}/\sqrt{N}+\ord(1/N)$, 
where we have used the overlap $\braket{s}{w}=1/\sqrt{N}$. 

Within this subspace, we find an effective Landau-Zener problem 
\begin{align}
\label{Landau-Zener}
\hat H(t)\to
-\Omega
\left(
\begin{array}{cc}
 1-f(t) & f(t)/\sqrt{N}
 \\ 
 f(t)/\sqrt{N} & f(t)
\end{array}
\right) 
+
\new{\ord\left(\frac1N\right)}
\,.
\end{align}
As a consequence, we obtain an avoided level crossing at $f(t)=1/2$ 
where the overlap $\braket{s}{w}=1/\sqrt{N}$ determines the minimum 
energy gap $\Delta E_{\rm min}=\Omega/\sqrt{N}$.
%

\begin{figure}
\begin{center}
\includegraphics[width=0.5\textwidth]{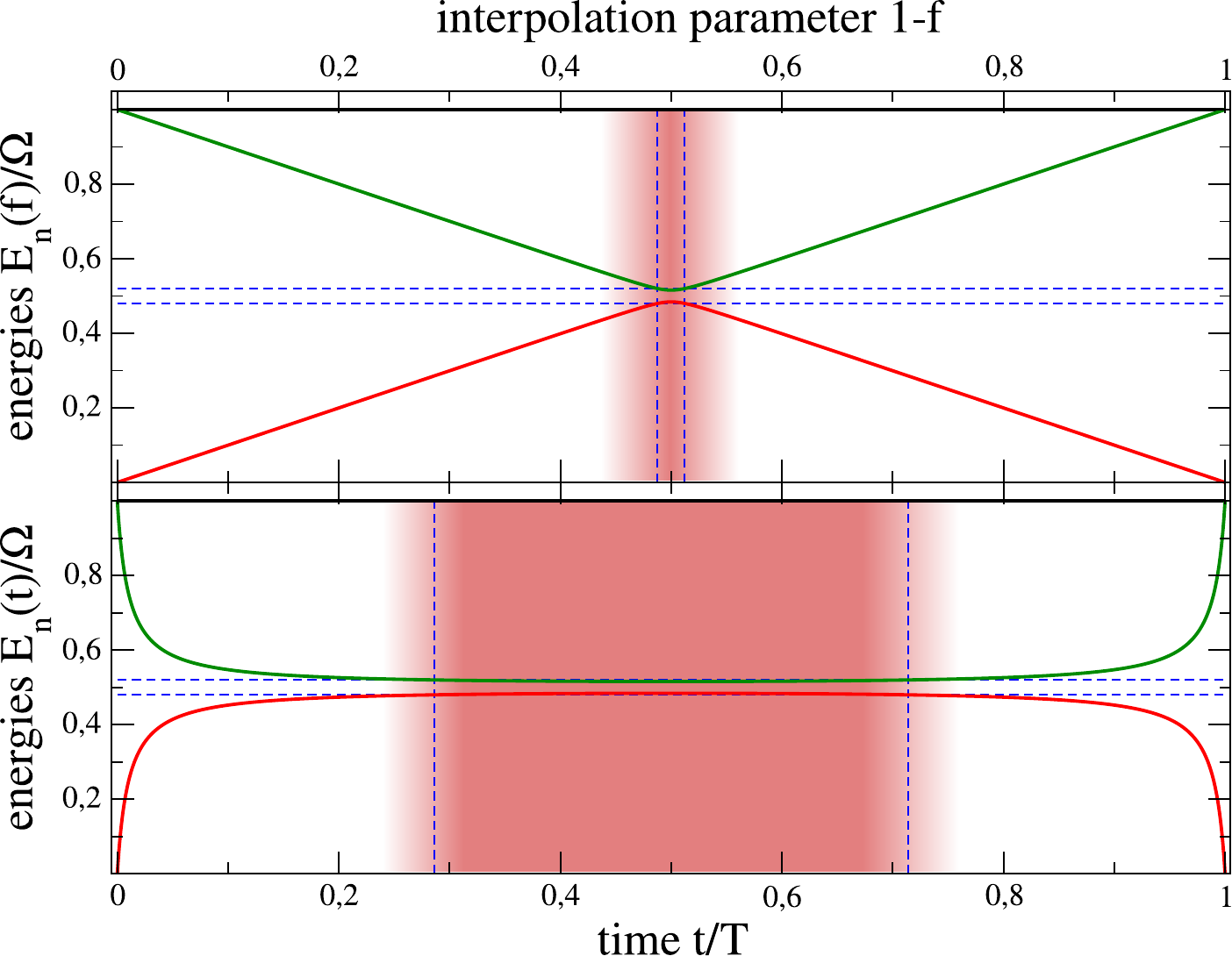}
\caption{\label{FIG:groverspectrum}
Top: Shifted spectrum of the Grover 
Hamiltonian~\eqref{EQ:adiabatic_search_algorithm} as a function of 
the interpolation parameter $f$ and $N=10^3$. 
For constant-speed interpolation $1-f(t) = t/T$, little time is spent 
in the region where the energy gap is small and thus the approximations 
usually employed for deriving a master equation become invalid (shaded region).
Bottom: Same spectrum as a function of time and for variable-speed interpolation $\dot f \propto \Delta E^2(t)$. 
The region where typical master equations become invalid is much larger.}
\end{center}
\end{figure}

Comparison with the adiabatic condition~\eqref{EQ:adiabatic} reveals that 
a constant speed interpolation $\dot f=\rm const$ requires a run-time 
$T$ of order $T=\ord(N)$ in order to satisfy~\eqref{EQ:adiabatic} and 
thus offers no quantum speed-up in comparison to a classical 
brute-force search. 
However, interpolations with adaptive speeds such as 
$\dot f(t)\propto[\Delta E(t)]^2$ can yield significantly shorter run-times 
$T$ of order $T=\ord(\sqrt{N})$ which is the usual quadratic quantum 
speed-up for the Grover search problem~\cite{roland2002a,schaller2006b}, 
see also Fig.~\ref{FIG:groverspectrum}.

\new{Quite generally, imperfections such as gate errors or the coupling 
to an environment can already deteriorate the quadratic quantum speed-up for the 
original Grover search problem, see, e.g., Refs.~\cite{long2000a,shapira2003a,shenvi2003a,regev2008a,ambainis2013a}.
In the following however, we focus on the adiabatic version of the 
Grover search problem.} 

\section{Perturbation Theory}\label{Perturbation Theory}

One of the motivations for considering adiabatic quantum algorithms was the robustness of the ground state against low-energy noise and decoherence~\cite{childs2001a,aberg2005a}. 
If we assume that the characteristic frequency \new{$\omega_{\rm env}$}  
and energy scales 
of the environment (such as its temperature) are much smaller than the initial 
and final energy gap $\Delta E_{\rm max}=\Omega$ of the 
Hamiltonian~\eqref{EQ:adiabatic_search_algorithm}, then one would expect that 
the unavoidable coupling to the environment does not cause problems for 
preparing the initial ground state $\ket{s}$ or reading out the final
ground state $\ket{w}$. 
However, since the energy gap becomes extremely small 
$\Delta E_{\rm min}=\Omega/\sqrt{N}$ at intermediate times, 
the impact of the environment may no longer be negligible
for the time evolution in between -- 
which is the question we want to study in the following\footnote{\new{
Of course, 
adiabatic quantum algorithms can also 
be affected by numerous error sources, such as non-adiabatic excitations, 
control errors or thermal excitations, see, e.g., Refs.~\cite{childs2001a,aberg2005a,jordan2006a,young2013a,sarovar2013a,asshab2014a,matsuura2017a,albash2017a,yan2022a}. 
}}. 

\new{In order approach this question, } 
we consider the joint Hilbert space of system plus environment 
with the total Hamiltonian 
\begin{align}
\label{total Hamiltonian}
\hat H=\hat H_{\rm sys}+\hat H_{\rm env}+\hat H_{\rm int}
\,,
\end{align}
where the system Hamiltonian is given by 
Eq.~\eqref{EQ:adiabatic_search_algorithm}
and acts on the system Hilbert space (i.e., the qubits), while the 
intrinsic dynamics of the environment is generated by the 
environment Hamiltonian $\hat H_{\rm env}$ which acts on the environment 
Hilbert space. 

The interaction between system and environment is governed by 
$\hat H_{\rm int}$ which acts on both Hilbert spaces. 
Neglecting correlated errors of two or more qubits~\cite{shor1995a,steane1996a,laflamme1996a,fowler2014a}, 
we start with the general ansatz~\footnote{\new{Correlated errors could be formally described by $\hat H_{\rm int} = g \Omega \sum_{\mu\nu} \sum_{ab} \hat \sigma^a_\mu \hat \sigma^b_\nu \hat B^{ab}_{\mu\nu}$ but would yield qualitatively similar results.}}
\begin{align}
\label{interaction}
\hat H_{\rm int}
=
g\Omega\sum_{\mu=1}^n \sum_a \hat\sigma_\mu^a \hat B_\mu^a 
\,,
\end{align}
where $\hat\sigma_\mu^a$ are the usual Pauli spin matrices acting on the 
qubit $\mu$ and $\hat B_\mu^a$ the corresponding bath operators acting 
on the environment.  
In order to have dimensionless operators $\hat\sigma_\mu^a$ and 
$\hat B_\mu^a$ of order unity, we inserted the energy pre-factor $\Omega$ 
and the small coupling strength $g\ll1$. 
\new{Naturally, this assumption $g\ll1$ limits our results to the 
weak-coupling regime. 
However, this is not a very strong limitation since well-isolated
quantum systems are typically required 
for quantum computation anyway 
(unless alternative concepts such as
quantum reservoir computing or dissipative quantum computing are invoked~\cite{verstraete2009b,kliesch2011a,sinayskiy2012a,fujii2021a,govia2021a}). 
For example, the assumption $g\ll1$ is typically employed for ensuring 
the straightforward preparation of the initial ground state and the 
eventual read-out of the final ground state.}

As a first approach, we employ a perturbative expansion in powers of 
this small coupling strength $g\ll1$. 
As usual, in view of the small coupling, 
we assume that the initial state factorizes (Born approximation) 
\begin{align} 
\label{factorizes}
\hat\rho(t_{\rm in})
=
\hat\rho_{\rm sys}(t_{\rm in})\otimes\hat\rho_{\rm env}^0
=
\hat\rho_{\rm sys}^{\rm in}\otimes\hat\rho_{\rm env}^0
\,,
\end{align}
%
and we employ the interaction picture where the unperturbed dynamics 
generated by $\hat H_{\rm sys}$ and $\hat H_{\rm env}$ apply to the 
operators $\hat\sigma_\mu^a$ and $\hat B_\mu^a$, respectively. 

\subsection{First Order}

Usually, the expectation value of the bath coupling operators is assumed to vanish \new{(e.g., if $\hat B_\mu^a$ is a linear combination of creation 
and annihilation operators while $\hat\rho_{\rm env}^0$ is a thermal state),}
but let us also consider the case where 
$\expval{\hat B_\mu^a}_{\rm env}^0\neq 0$.
Then, to first order in $g$, the state of the system evolves 
\new{from $t_{\rm in}$ to $t_{\rm out}$ (to be specified below)} 
as 
%
%
\begin{align}
\label{first-order}
\hat\rho_{\rm sys}^{\rm out}
&=
\hat\rho_{\rm sys}^{\rm in}
-i\int\limits_{t_{\rm in}}^{t_{\rm out}}dt\,
\left[\langle\hat H_{\rm int}(t)\rangle_{\rm env}^0, 
\hat\rho_{\rm sys}^{\rm in}\right]
+\ord(g^2)
\,,
\end{align}
where $\langle\hat H_{\rm int}(t)\rangle_{\rm env}^0
={\rm Tr}_{\rm env}\{\hat H_{\rm int}(t)\hat\rho_{\rm env}^0\}$
denotes the average over the environment degrees of freedom. 
Up to this level, we find that the impact of the environment 
leads to an effective shift 
$\hat H_{\rm sys}\to\hat H_{\rm sys}^{\rm eff}=\hat H_{\rm sys}+
\langle\hat H_{\rm int}\rangle_{\rm env}^0$
of the system Hamiltonian. 
To study this shift, we may insert the identity operator in the 
system Hilbert space
\begin{align}
\label{identity}
\f{1}_{\rm sys}
&=
\ket{w}\bra{w}+\ket{w_\perp}\bra{w_\perp}
+\sum_{\rm rest}\ket{\rm rest}\bra{\rm rest}
\nn
&=
\ket{w}\bra{w}+\ket{s}\bra{s}
-\frac{\ket{w}\bra{s}+\ket{s}\bra{w}}{\sqrt{N}}
+
\new{ 
\sum_{\rm rest}\ket{\rm rest}\bra{\rm rest}
+\ord\left(\frac1N\right)
}
\,,
\end{align}
where the states $\ket{\rm rest}$ are orthogonal to the sub-space spanned 
by $\ket{w}$ and $\ket{w_\perp}$. 
The resulting modification of the Landau-Zener Hamiltonian~\eqref{Landau-Zener}
is then determined by the matrix elements of the Pauli spin operators
$\hat\sigma_\mu^a$ multiplied by the expectation values 
$\langle\hat B_\mu^a\rangle_{\rm env}^0$ of the bath operators. 
Only the diagonal matrix elements $\bra{w}\hat\sigma_\mu^z\ket{w}=\pm1$  
and $\bra{s}\hat\sigma_\mu^x\ket{s}=1$ are of order unity, the rest is zero 
or suppressed as $\ord(1/\sqrt{N})$, e.g., 
$\bra{w}\hat\sigma_\mu^z\ket{s}=\pm1/\sqrt{N}$ or 
$\bra{w}\hat\sigma_\mu^x\ket{s}=1/\sqrt{N}$. 

As a consequence, the off-diagonal elements of the 
Landau-Zener Hamiltonian~\eqref{Landau-Zener}
acquire corrections of order $\ord(g/\sqrt{N})$ 
which do not have a significant effect for $g\ll1$. 
However, the corrections to the diagonal elements of the 
Landau-Zener Hamiltonian~\eqref{Landau-Zener} are much less 
strongly suppressed and scale with $\ord(g)$.
Even though they do \new{basically} 
not change the size of the minimum gap 
\new{$\Delta E_{\rm min}=\Omega/\sqrt{N}+\ord(g\Omega/\sqrt{N})$,}
they may shift the position of the minimum gap a bit 
because the environment induced energy shift for the state 
$\ket{w}$ may well be different from the shift of the other 
state $\ket{s}$ if $\langle\hat B_\mu^z\rangle_{\rm env}^0\neq
\langle\hat B_\mu^x\rangle_{\rm env}^0$. 
If these expectation values are unknown, this
could be problematic for executing 
interpolations with adaptive speeds 
such as $\dot f(t)\propto[\Delta E(t)]^2$, which were required for 
reaching the quantum speed-up of the Grover search algorithm. 

Note that there are also non-zero matrix elements of the 
Pauli spin operators $\hat\sigma_\mu^a$ between the states 
$\ket{w}$ or $\ket{s}$ on the one hand and the remaining states 
$\ket{\rm rest}$ on the other hand. 
As a result, the eigen-states of the shifted system Hamiltonian
$\hat H_{\rm sys}^{\rm eff}$ may also contain a small 
\new{$\ord(g)$} 
admixture 
of those states $\ket{\rm rest}$ but this does not affect the 
main properties of the Landau-Zener problem~\eqref{Landau-Zener}. 

\subsection{Second Order}

As we have observed above, the first-order result~\eqref{first-order} 
amounts to a shift of the system Hamiltonian
$\hat H_{\rm sys}\to\hat H_{\rm sys}^{\rm eff}=\hat H_{\rm sys}+
\expval{\hat H_{\rm int}}_{\rm env}^0$.
%
If necessary, we may accordingly renormalize the system and interaction Hamiltonians by $\hat H_{\rm sys}\to\hat H_{\rm sys}^{\rm eff}$ and $\hat H_{\rm int}\to\hat H_{\rm int}^{\rm ren} = 
\hat H_{\rm int}-\expval{\hat H_{\rm int}}_{\rm env}^0$, after which
we have $\langle\hat H_{\rm int}^{\rm ren}\rangle_{\rm env}^0=0$ 
and thus the lowest non-trivial contribution is the second order (and the interaction picture for the system operators is then understood with 
respect to $\hat H_{\rm sys}^{\rm eff}$)
\begin{align}
\label{second-order}
\hat\rho_{\rm sys}^{\rm out}
&=
\hat\rho_{\rm sys}^{\rm in}
+
g^2\Omega^2 
\sum_{\mu\nu ab}
\int\limits_{t_{\rm in}}^{t_{\rm out}}dt_1 
\int\limits_{t_{\rm in}}^{t_1}dt_2 
\Big[ 
\langle\hat B_\mu^a(t_1)\hat B_\nu^b(t_2)\rangle_{\rm env}^0
\left(
\hat\sigma_\mu^a(t_1) \hat\rho_{\rm sys}^{\rm in} \hat\sigma_\nu^b(t_2)
-
\hat\sigma_\mu^a(t_1) \hat\sigma_\nu^b(t_2) \hat\rho_{\rm sys}^{\rm in}
\right)
+{\rm h.c.} 
\Big] 
\nn
&\qquad
\new{+\ord(g^3)}
\,.
\end{align}
Assuming adiabatic evolution~\eqref{EQ:adiabatic} for the system
$\hat H_{\rm sys}$,
the error probability $P_{\rm error}$
(i.e., the probability for not ending up in the final ground state of 
$\hat H_{\rm sys}$)  
is caused by the interaction with the environment and corresponds to 
the $\ord(g^2)$ corrections in Eq.~\eqref{second-order}. 
Given a quite general environment, let us now estimate the order of magnitude 
of the error probability $P_{\rm error}$.

For a stationary environment, the correlation functions 
$\langle\hat B_\mu^a(t_1)\hat B_\nu^b(t_2)\rangle_{\rm env}^0$
depend on the time difference $t_1-t_2$ only.
Here we assume that they decay on a characteristic time scale 
$\tau_{\rm env}$ and may feature oscillations with typical frequency 
scales $\omega_{\rm env}$.
It is advantageous to distinguish between positive 
and negative environment frequencies 
\new{$\omega_{\rm env}$ as arguments of the Fourier transform of 
$\langle\hat B_\mu^a(t)\hat B_\nu^b(0)\rangle_{\rm env}^0$
which is proportional to the reservoirs spectral coupling density.
One can take the convention that the}
former $\omega_{\rm env}>0$ mediate excitations of the system 
while the latter $\omega_{\rm env}<0$ correspond to de-excitations. 
In order to avoid unwanted excitations from the initial and final ground states 
(as motivated in the Introduction), the positive environment frequencies 
$\omega_{\rm env}>0$ should be much smaller than the initial and final gap 
$\omega_{\rm env}\ll\Omega$.
For a thermal environment, this means that the environment temperature 
should also be far below $\Omega$ (unless the environment features a large 
gap in its density of states or spectral coupling density). 

Transforming the double time integration in Eq.~\eqref{second-order} 
into the relative $t_-=t_1-t_2$ and the ``centre-of-mass'' coordinate  
$t_+=(t_1+t_2)/2$ allows to estimate the scaling:
As the interaction picture operators $\hat \sigma^a_\mu(t)$ remain normalized,
the $t_-$ integration scales with $\tau_{\rm env}$.
Additional problems could arise for a highly non-Markovian 
environment where the spectral function 
is not well behaved in the infra-red~\cite{haenggi1978a,tiersch2007a,jin2016a} such that 
$\langle\hat B_\mu^a(t_1)\hat B_\nu^b(t_2)\rangle_{\rm env}^0$ 
decays very slowly as a function of $t_1-t_2$, but we do not consider 
this case here. 
Assuming that the total integration time $\Delta t=t_{\rm out}-t_{\rm in}$ 
is much larger than $\tau_{\rm env}$, the remaining time integral over 
$t_+$ scales with $\Delta t$. 

Finally, depending on the spatial range of the correlations in the reservoir, 
the double sum over all qubits $\sum_{\mu,\nu}$ scales with their number $n$
for short-range correlations or with their number squared $n^2$ for 
long-range correlations.
Taking the latter case as an upper bound, we may estimate the order of magnitude of the error probability via 
\begin{align}
\label{estimate}
P_{\rm error}\leq\ord\left(g^2\Omega^2n^2\tau_{\rm env}\Delta t\right) 
\,.
\end{align}
Unfortunately, a small coupling $g\ll1$ to the environment might not be 
sufficient to ensure a small error probability.
As explained in Sec.~\ref{Adiabatic Quantum Search Algorithm},
the total run-time $T$ scales with $T=\ord(\sqrt{N}/\Omega)$ 
and thus $\Delta t$ can be very large, i.e., grow exponentially with 
the number $n$ of qubits. 
Since it is probably unrealistic to assume that the coupling $g$ 
of each qubit to the environment 
(or the environment correlation time $\tau_{\rm env}$)
decreases exponentially if we increase 
the number $n$ of qubits, i.e., the system size, the above 
estimate~\eqref{estimate} does not allow us to conclude a small 
error probability. 

Let us consider the integrand in Eq.~\eqref{second-order} in more detail. 
Since the coupling $g\ll1$ is small and $N\ggg1$ is exponentially large, 
we neglect terms of order $\ord(g^2/\sqrt{N})$ in the following. 
Then assuming adiabatic evolution~\eqref{EQ:adiabatic} for the system 
$\hat H_{\rm sys}$, the unitary system dynamics with 
$\hat\sigma_\mu^a(t)=
\hat U_{\rm sys}^\dagger(t)\,\hat\sigma_\mu^a\,\hat U_{\rm sys}(t)$
can be approximated by~\footnote{Note that the geometrical phase vanishes for the adiabatic Grover search.}
\begin{align}
\label{unitary}
\hat U_{\rm sys}(t)
&=
\ket{\psi_0(t)}
e^{-i\varphi_0^{\rm dyn}(t)}
\bra{s}
+
\ket{\psi_1(t)}
e^{-i\varphi_1^{\rm dyn}(t)}
\bra{w}
+
\new{ 
\sum_{\rm rest}\ket{\rm rest}\bra{\rm rest}
+\ord\left(\frac{1}{\sqrt{N}}\right)
} 
\,.
\end{align}
While $\hat U_{\rm sys}(t)$ does not affect the states $\ket{\rm rest}$ 
orthogonal to $\ket{s}$ and $\ket{w}$, it acts as a rotation with a 
time-dependent angle $\varphi(t)$ within this sub-space. 
More precisely, up to the dynamical phases
$\dot\varphi_n^{\rm dyn}(t)=E_n(t)$,
the transformation~\eqref{unitary} describes a rotation from the initial ground 
$\ket{\psi_0(t=0)}=\ket{s}$ and first excited states 
$\ket{\psi_1(t=0)}=\ket{s_\perp}\approx\ket{w}$ to the instantaneous ground 
$\ket{\psi_0(t)}$ and first excited state $\ket{\psi_1(t)}$. 

Even though the matrix elements of the Pauli spin matrices $\sigma_\mu^a$
between the states $\ket{s}$ and $\ket{w}$ are exponentially suppressed,
this is no longer true for the matrix elements between the
instantaneous ground and the first excited state, e.g.,
$\bra{\psi_0(t)}\sigma_\mu^z\ket{\psi_1(t)}=
\pm\sin[\varphi(t)]\cos[\varphi(t)]$.
Since the rotation angle $\varphi(t)$ changes from zero to $\pi$ when traversing the Landau-Zener transition, the matrix elements mainly
contribute in the vicinity of the minimum gap -- as one might expect.

In principle, there are also non-zero matrix elements between the
instantaneous eigen-vectors
$\ket{\psi_0(t)}$ and $\ket{\psi_1(t)}$ on the one hand and the
remaining vectors $\ket{\rm rest}$ on the other hand.
However, since the dynamical phases
$\dot\varphi_0^{\rm dyn}(t)=E_0(t)$ and
$\dot\varphi_1^{\rm dyn}(t)=E_1(t)$ are rapidly oscillating
with a frequency $\ord(\Omega)$, while the positive environment
frequencies $\omega_{\rm env}\ll\Omega$ are much smaller,
the time integrals in Eq.~\eqref{second-order} become
strongly suppressed for those transitions and thus can be neglected.

For the same reason, transitions from $\ket{\psi_0(t)}$ to $\ket{\psi_1(t)}$ 
occur predominantly in the vicinity of the minimum gap where the rapid 
oscillations
$\dot\varphi_0^{\rm dyn}(t)$ and
$\dot\varphi_1^{\rm dyn}(t)$ almost cancel each other.

\section{Master Equation}\label{Master Equation} 

The fact that the error probability~\eqref{estimate} can become large 
for long enough time intervals $\Delta t$ shows that one should not 
apply perturbation theory~\eqref{second-order} to the whole run-time $T$ 
which is exponentially long $T=\ord(\sqrt{N}/\Omega)$.
Thus, in order to understand the long-time behavior, a different method 
is required. 
One option is a suitable master equation, see, e.g.,~\cite{weiss1993,breuer2007,schlosshauer2007}.
Master equations are usually derived 
(e.g., in the Nakajima-Zwanzig approach~\cite{nakajima1958a,zwanzig1960a})
via a series of approximations, 
such as Born-Markov and secular approximations. 
The Born-Markov approximation can be motivated by assuming that the 
coupling $g$ is small and the environment is large such that all 
effects on the environment (e.g., correlations between system and 
environment) are quickly dissipated in the environment, i.e., 
transported away from the system. 
As a result, the system always ``sees'' the same environment, 
i.e., the environment has no memory, and we can use the
ansatz~\eqref{factorizes} for each time step.
In the discussion based on perturbation theory~\eqref{second-order}
above, this is related to the condition that the environment correlation
time $\tau_{\rm env}$ is much smaller than $\Delta t$. 

The secular approximation, on the other hand, can be motivated by 
assuming that the phase differences between the instantaneous 
eigen-states of $\hat H_{\rm sys}$ are rapidly oscillating such 
that their time integrals can be neglected. 
While this assumption is justified for the phase difference between 
the instantaneous ground state $\ket{\psi_0(t)}$ and the remaining 
states $\ket{\rm rest}$, for example, it is not justified for the 
phase difference between the instantaneous ground state $\ket{\psi_0(t)}$
and the first excited state $\ket{\psi_1(t)}$ which changes extremely 
slowly in the vicinity of the minimum gap. 
Thus, the usual secular approximation \new{in the system energy eigen-basis} is problematic for the scenario 
of interest here~\cite{schaller2008a,majenz2013a,maekelae2013a,matus2023a}.

\new{This failure of the secular approximation (at least in its usual from) 
for small system gaps has already been discussed in a number of works, 
e.g., for stationary problems (see, e.g., Refs.~\cite{maekelae2013a,hartmann2020a,trushechkin2021a}) or for dissipative 
Landau-Zener sweeps (see, e.g., Refs.~\cite{saito2007a,zueco2008a,nalbach2009a,kahlert2024a,dai2025a}).} 

\subsection{Coarse Grained Master Equation} 
\label{Coarse Grained Master Equation} 

In order to overcome this obstacle and to obtain a master equation without 
the secular approximation, we employ a coarse graining procedure, see also~\cite{schaller2008a,benatti2009a,majenz2013a,rivas2017a}.
For simplicity, we only assume coupling terms~\eqref{interaction} in  
$\sigma_\mu^z$ direction for now, other coupling terms will be 
discussed below. 

We split the total evolution time $T$ into many time steps 
$\Delta t$. 
They should be much larger than the correlation time $\tau_{\rm env}$
of the environment and, as discussed above, we apply the 
Born approximation~\eqref{factorizes} at each time step.
However, these time intervals $\Delta t$ should still be small enough 
that we may apply perturbation theory~\eqref{second-order} 
for each time step. 
This is possible because the coupling $g$ is small. 
Furthermore, $\Delta t$ should be much smaller than the total run-time $T$
such that the change of $\hat U_{\rm sys}(t)$ during $\Delta t$ can be neglected.

In this regime, we may apply basically the same arguments as after 
Eq.~\eqref{second-order} to obtain the form 
\begin{align}
\label{coarse-grained} 
\hat\rho_{\rm sys}(t+\Delta t)
&\approx
\hat\rho_{\rm sys}(t)
-\ii \new{\Delta t}\left[\Delta \hat H^{\rm eff}_{\rm sys}, \hat \rho_{\rm sys}(t)\right]
+
\new{ 
\Delta t\Big[
\Big(
\hat L\hat\rho_{\rm sys}(t)\hat L-
\hat L^2
\hat\rho_{\rm sys}(t)
\Big)
+{\rm h.c.}
\Big]
}
\,,
\end{align}
where $\Delta \hat H^{\rm eff}_{\rm sys}=\ord(g^2)$ is a small Lamb-shift type level renormalization term.
In the sub-space spanned by $\ket{s}$ and $\ket{w}$, we may use 
$\hat U_{\rm sys}^\dagger(t)
\sigma_\mu^z
\hat U_{\rm sys}(t)
\approx 
\hat U_{\rm sys}^\dagger(t)
\ket{w}\bra{w}
\hat U_{\rm sys}(t)$
from Eq.~\eqref{unitary}, such that 
we find the effective Lindblad operator 
\begin{align}
\label{Lindblad-operator}
\hat L(t)
=
\sqrt{\Gamma}\,
\hat U_{\rm sys}^\dagger(t)
\ket{w}\bra{w}
\hat U_{\rm sys}(t)
=
\hat L^\dagger(t)
\,,
\end{align}
which is self-adjoint and whose strength is governed by the effective 
decay rate, cf.~Eq.~\eqref{estimate}
\begin{align}
\label{Gamma-Coarse}
\Gamma=\ord(g^2\Omega^2n^2\tau_{\rm env})
\,.
\end{align}
Transforming from the interaction picture to the Schr\"odinger
picture, we get the same form as in the Lindblad master
equation~\eqref{EQ:master} 
\begin{align}
\label{coarse}
\frac{d\hat\rho_{\rm sys}}{dt}=-\ii\left[\hat H_{\rm sys}^{\rm eff},\hat\rho\right]+
\hat L\hat\rho\hat L^\dagger-
\frac12\left\{\hat L^\dagger\hat L,\hat\rho\right\} 
\end{align}
with the effective system Hamiltonian
$\hat H_{\rm sys}^{\rm eff}=\hat H_{\rm sys}+
\Delta \hat H^{\rm eff}_{\rm sys}$
while the Lindblad operator $\hat L$
is given by $\hat L=\ket{w}\sqrt{\Gamma}\bra{w}$.
This dissipator destroys coherences between the states $\ket{s}$ 
and $\ket{w}$ and thus slows down transitions in complete analogy 
to the quantum Zeno effect. 
As an intuitive picture, the environment is permanently
measuring whether the system is in the state $\ket{w}$
or not. 
Since the effective measurement frequency $\Gamma$ is not
exponentially suppressed (it even increases polynomially with $n$)
while the transition rate governed by $\hat H_{\rm sys}$
is exponentially small \new{$\omega_{\rm sys}=\Omega/\sqrt{N}$}
in the vicinity of the minimum gap, 
we are in the regime of the quantum Zeno effect for large 
enough $n$. 

\new{Note that the condition $\omega_{\rm sys}\ll\Gamma$ for being in the 
quantum Zeno regime is exactly opposite to that $\omega_{\rm sys}\gg\Gamma$
typically used for justifying the secular approximation~\cite{breuer2007,trushechkin2021a}. 
As a result, studying the robustness of adiabatic quantum algorithms 
by using the usual master equation based on the secular approximation
becomes unreliable in the limit of large $n$ considered here, where one 
would enter the quantum Zeno regime.}

\subsection{Redfield Equation}\label{Redfield Equation} 

As an alternative approach, we may start from the Redfield-I equation~\cite{redfield1965a}.
This equation is also based on the Born-Markov approximation, 
but does not employ the secular approximation. 
Specifying it to our set-up in Eqns.~\eqref{total Hamiltonian} 
and~\eqref{interaction} in the interaction picture
and again assuming coupling 
terms~\eqref{interaction} in $\sigma_\mu^z$ direction only, 
the Redfield-I equation reads 
\begin{align}
\label{Redfield}
\frac{d\hat\rho_{\rm sys}(t)}{dt}
&=
g^2\Omega^2 
\sum_{\mu\nu}\int\limits_0^t dt'\, 
\langle\hat B_\mu^z(t)\hat B_\nu^z(t')\rangle_{\rm env}^0
\nn
&\times 
\Big(
\hat\sigma_\nu^z(t') \hat\rho_{\rm sys}(t) \hat\sigma_\mu^z(t)
-
\frac12 
\left\{
\hat\sigma_\mu^z(t)\hat\sigma_\nu^z(t'),\hat\rho_{\rm sys}(t)
\right\}
+\frac12 
\left[
\hat\rho_{\rm sys}(t),\hat\sigma_\mu^z(t)\hat\sigma_\nu^z(t')
\right]
\Big)
\nn
&+g^2\Omega^2 
\sum_{\mu\nu}\int\limits_0^t dt'\, 
\langle\hat B_\mu^z(t')\hat B_\nu^z(t)\rangle_{\rm env}^0
\nn
&\times 
\Big(
\hat\sigma_\nu^z(t) \hat\rho_{\rm sys}(t)\hat\sigma_\mu^z(t')
-
\frac12 
\left\{
\hat\sigma_\mu^z(t')\hat\sigma_\nu^z(t),\hat\rho_{\rm sys}(t)
\right\}
+\frac12 
\left[
\hat\rho_{\rm sys}(t),\hat\sigma_\mu^z(t')\hat\sigma_\nu^z(t)
\right]
\Big)
\,.
\end{align}
If we use the same assumption as before, i.e., that the environment 
correlation time $\tau_{\rm env}$ is much shorter than the 
(exponentially long) time scale on which the system dynamics changes, 
we may approximate 
$\hat\sigma_\mu^z(t')\approx\hat\sigma_\mu^z(t)$ which allows us to 
perform the $t'$ integration over the environment correlation 
function
\begin{align} 
\label{Gamma-Redfield}
\Gamma
=
g^2\Omega^2 
\sum_{\mu\nu}\int\limits_0^t dt'\, 
\langle\hat B_\mu^z(t)\hat B_\nu^z(t')\rangle_{\rm env}^0
\,.
\end{align}
As a result, we find the same Lindblad master equation~\eqref{coarse} 
\new{with} a small additional shift of the system Hamiltonian due to 
the commutator terms 
$[\hat\rho_{\rm sys}(t),\hat\sigma_\mu^z(t)\hat\sigma_\nu^z(t)]$
in the \new{second and fourth} 
line of Eq.~\eqref{Redfield}. 

Including the $\sigma_\mu^x$ coupling to the environment and using 
$ \hat U_{\rm sys}^\dagger(t)
\sigma_\mu^x
\hat U_{\rm sys}(t)
\approx 
\hat U_{\rm sys}^\dagger(t)
\ket{s}\bra{s}
\hat U_{\rm sys}(t)$
would give an additional Lindblad operator $\ket{s}\bra{s}$ and thus even
enhance the quantum Zeno effect, see App.~\ref{SEC:lindblad_slow}.

Either way, the quantum Zeno effect would negate the performance of 
the adiabatic Grover search algorithm since the run-time 
$T=\ord(\sqrt{N})$ describing the usual quadratic quantum speed-up 
for the Grover search problem in the absence of any environment 
does not work anymore.
As an intuitive picture, the environment permanently measures whether 
the system is still in the state $\ket{s}$ or already in the state 
$\ket{w}$ and these permanent measurements inhibit transitions from 
$\ket{s}$ to $\ket{w}$. 
From another perspective, these effective measurements destroy the 
coherences between the states $\ket{s}$ and $\ket{w}$ such that 
we can no longer add up amplitudes, but we are back to adding up 
probabilities -- such that the main quantum advantage is gone.

In analogy to Sec.~\ref{Quantum Zeno Effect}, one could employ a
much longer run-time after which the state approaches the steady 
state $\hat\rho=\f{1}/2$ (in the Landau-Zener sub-space), 
but this enlarged run-time would scale with
$\ord(N\Gamma/\Omega^2)$ which means that we lost the
quadratic quantum speed-up for the Grover search problem
and come back to the usual linear scaling $\ord(N)$
of a classical brute-force search. 


\section{Generalization}\label{Generalization} 

Having found that the quantum Zeno effect can pose strong restrictions on 
the performance of the adiabatic quantum search algorithm considered above, 
one might object that this could be an artifact of the special choice of 
the Hamiltonian~\eqref{EQ:adiabatic_search_algorithm} which would be absent 
for other Hamiltonians.
\new{In the following, we discuss how to generalize our findings to 
generic Hamiltonians and show that analogous restrictions should apply
under suitable conditions.} 

Unfortunately, \new{this generalization} 
is complicated by the fact that the level structure of general adiabatic 
quantum algorithms is not fully understood.
If we use the same initial projector Hamiltonian as in 
Eq.~\eqref{EQ:adiabatic_search_algorithm} but replace the 
final Hamiltonian by a more realistic version
(e.g., bi-linear or tri-linear polynomials of $\hat\sigma_\mu^z$ 
matrices), we can still 
obtain analytic \new{estimates of}
the eigen-states or eigen-energies and 
it has been shown that the minimum gap does also scale exponentially 
as $1/\sqrt{N}$ in this case~\cite{znidaric2006a}. 
Going one step further and also replacing the initial projector 
Hamiltonian by a paramagnetic Hamiltonian $\propto\sum\sigma^x_\mu$
with the same initial ground state $\ket{s}$, for example, 
it has been speculated~\cite{farhi2001a} that it might be possible to 
achieve 
polynomial scaling of the minimum gap for NP-complete problems~\cite{lucas2014a}, 
but this speculation is under debate. 
For example, it has been argued~\cite{altshuler2010a} that disorder induced localization 
phenomena should induce a scaling behavior which is even worse than 
the Grover scaling, i.e., that the minimum gap shrinks faster than 
$1/\sqrt{N}$ for large $N$. 
In this context, it should also be mentioned that the polynomial 
equivalence between adiabatic and sequential quantum computing 
has been shown in~\cite{aharonov2007a,mizel2007a}.
This implies that adiabatic quantum algorithms 
(in the absence of any environment) can achieve an
exponential speed-up for factoring -- but it does not necessarily 
imply a similar speed-up for NP-complete problems. 
At present, it is probably fair to say that this issue has not been 
fully settled yet. 

Nevertheless, we can still draw some conclusions based on general 
arguments. 
First of all, adiabatic quantum algorithms are presumably not able 
to perform non-trivial quantum computations ``for free''.  
Hence, 
one would expect that the (instantaneous) energy gap between the 
ground state and the first excited state(s) becomes very small at 
some point.
The external variation, such as $f(t)$ in 
Eq.~\eqref{linear-interpolation}, must be very slow in order to stay 
in the adiabatic regime.  
This point displays similarities to the critical point in a 
quantum phase transition and a major part of the quantum computation,
i.e., a strong change of the quantum (ground) state, is expected to 
occur in the vicinity of this point~\cite{latorre2004a,schuetzhold2006b,joerg2008a,amin2009a,joerg2010a,schaller2010a,ostilli2022a,ostilli2023a}.

In order to support this expectation, let us consider the 
instantaneous eigen-states 
$\hat H(t)\ket{\psi_n(t)}=E_n(t)\ket{\psi_n(t)}$ 
where $\ket{\psi_n(t)}$ denotes the parametric time dependence, 
which should not be confused with the dynamical time dependence
of the state $\ket{\psi(t)}$ as a solution of the Schr\"odinger 
equation. 
In the adiabatic regime, they are approximately the same, up to the 
phases mentioned in Sec.~\ref{Adiabatic Quantum Algorithms}. 
Now let us take the parametric time derivative 
$\ket{\partial_t\psi_n}$ and split it up into a parallel part 
$\ket{\partial_t\psi_n}^\|\propto\ket{\psi_n}$
(which can be eliminated by a suitable phase transformation) 
and the remaining perpendicular part $\ket{\partial_t\psi_n}^\perp$ 
which describes a real rotation in the Hilbert space.  
Taking the (parametric) time derivative of the equation 
$\hat H(t)\ket{\psi_n(t)}=E_n(t)\ket{\psi_n(t)}$, we find 
\begin{align}
\label{paramagnetic}
\left|\ket{\partial_t\psi_0}^\perp\right|^2
\leq
\frac{\bra{\psi_0}(\partial_t\hat H)^2
\ket{\psi_0}}{\Delta E_0^2}
\,,
\end{align}
where $\Delta E_0$ denotes the (instantaneous) energy gap between the 
ground state and the first excited state(s). 
\new{Even though Eq.~\eqref{paramagnetic} is only an upper bound, it 
supports the expectation that the major part of the quantum computation, 
i.e., the strongest change of the quantum (ground) state, 
is linked to the minimum (or minima) of the energy gap $\Delta E_0$, 
as in the Grover search algorithm~\eqref{EQ:adiabatic_search_algorithm}.} 

As an intuitive picture, one can visualize $\hat H(t)$ via a 
slowly changing effective potential landscape which has many 
local minima (for non-trivial problems to be solved). 
Starting in the initial ground state, i.e., the global minimum, 
and deforming the potential landscape, another local minimum 
\new{somewhere else} 
may become energetically favorable at some point (the critical point) 
and the system has to make a (tunneling) transition from the 
old to the new global minimum. 
\new{This scenario would be analogous to a first-order transition.}
As in the Landau-Zener problem~\eqref{Landau-Zener}, the 
(tunneling) transition rate determines the minimum gap $\Delta E_0$
at that point. 

Now, in the absence of symmetries (see~\cite{schuetzhold2006b,schaller2010a}), one would naturally 
expect that there are probably not more than two competing local 
minima at that critical point. 
As a result -- and because the minimum gap $\Delta E_0$ is very small -- 
it should be a good approximation to focus on these two states 
(representing the two competing local minima) for understanding the 
(tunneling) transition. 
Indeed, many of the adiabatic quantum algorithms discussed 
in the literature can be reasonably well approximated by more or less 
isolated Landau-Zener like avoided level crossings between the two 
lowest (instantaneous) energy eigen-states 
(even though there can be many more avoided level crossings at 
higher lying states). 

\subsection{Generalized Landau-Zener Transition}
\label{Generalized Landau-Zener Transition}

As motivated above, let us consider a general adiabatic quantum algorithm 
featuring a single isolated Landau-Zener like avoided level crossing which 
mediates the transition between the two states $\ket{\rm in}$ and 
$\ket{\rm out}$. 
Here  $\ket{\rm in}$ and $\ket{\rm out}$ denote the ground states just 
before and just after the avoided level crossing.
They are not necessarily the initial and final ground states, because 
the ground state could also change away from the avoided level crossing, 
see Eq.~\eqref{paramagnetic}. 

However, as the transition from $\ket{\rm in}$ to $\ket{\rm out}$
is supposed to constitute a major part of the quantum computation,
the two states are assumed to be macroscopically different.  
Here macroscopically different means that $\ket{\rm in}$ and 
$\ket{\rm out}$ disagree for at least a large number $\ord(n)$ of qubits. 
In the Grover search algorithm~\eqref{EQ:adiabatic_search_algorithm}, 
we have  
$\ket{\rm in}=\ket{s}=
\ket{\rightarrow}\ket{\rightarrow}\dots\ket{\rightarrow}$
and 
$\ket{\rm out}=\ket{w}=\ket{01\dots10}=
\ket{\downarrow}\ket{\uparrow}\dots
\ket{\uparrow}\ket{\downarrow}$, for example.  
As a result, the overlap between $\ket{\rm in}$ and $\ket{\rm out}$ 
is exponentially small 
$\braket{\rm in}{\rm out}=\exp\{-\ord(n)\}$, e.g., 
$\braket{s}{w}=1/\sqrt{N}$.

In this case, the matrix elements of the Pauli spin operators 
$\bra{\rm in}\hat\sigma_\mu^a\ket{\rm out}=\exp\{-\ord(n)\}$
are also exponentially small -- and the same applies to products 
of a finite number of Pauli spin operators, such as 
$\hat\sigma_\mu^a\hat\sigma_\nu^b$. 
If we assume a local Hamiltonian, i.e., a sum of polynomially many terms, 
each acting on a limited number of qubits (e.g., up to two or three), 
the off-diagonal (transition) matrix element is also exponentially small 
\new{$\bra{\rm in}\hat H(t)\ket{\rm out}=\exp\{-\ord(n)\}$.}

The Landau-Zener problem~\eqref{Landau-Zener} 
then generalizes to~\footnote{\new{Strictly speaking, in analogy to 
Eq.~\eqref{Landau-Zener}, the orthogonal basis vectors in this subspace 
can be chosen as $\ket{\rm in}$ and 
$\ket{\rm in}_\perp\propto(\ket{\rm out}-
\ket{\rm in}\braket{\rm in}{\rm out})$. 
However, as $\ket{\rm in}$ and $\ket{\rm out}$ are nearly orthogonal, 
this would not significantly change the outcome. 
}} 
\begin{align}
\label{Landau-Zener-general}
\hat H\new{(t)}\to
\left(
\begin{array}{cc}
\bra{\rm in}\hat H\new{(t)}\ket{\rm in} & \bra{\rm in}\hat H\new{(t)}\ket{\rm out}
 \\ 
\bra{\rm out}\hat H\new{(t)}\ket{\rm in} & \bra{\rm out}\hat H\new{(t)}\ket{\rm out}
\end{array}
\right) 
\,,
\end{align}
and thus the minimum gap $\Delta E_{\rm min}$ 
occurring when the diagonal elements are equal 
\new{or at least extremely close}
$\bra{\rm in}\hat H\new{(t)}\ket{\rm in}\new{\approx}\bra{\rm out}\hat H\new{(t)}\ket{\rm out}$
is determined by the off-diagonal matrix element $\Delta E_{\rm min}\new{\approx {\rm min}\{2|\bra{\rm in}\hat H(t)\ket{\rm out}|\}}$.  
Consequently, $\Delta E_{\rm min}$ is also exponentially small 
$\Delta E_{\rm min}=\exp\{-\ord(n)\}$ which means that achieving an 
exponential speed-up for such an adiabatic quantum algorithm 
would require violating some of the assumptions above.

Coupling this more general system~\eqref{Landau-Zener-general}
to an environment as in Eqns.~\eqref{total Hamiltonian} 
and \eqref{interaction}, we may now go through basically the same 
steps as in Secs.~\ref{Perturbation Theory} and \ref{Master Equation}. 
The unitary system dynamics governing the time evolution in the 
vicinity of the avoided level crossing can be approximated as 
in Eq.~\eqref{unitary}, but with $\ket{s}$ and $\ket{w}$ 
being replaced by $\ket{\rm in}$ and $\ket{\rm out}$, respectively. 
Deriving the effective Lindblad operators via the coarse grained 
master equation as in Sec.~\ref{Coarse Grained Master Equation} 
or the Redfield equation in Sec.~\ref{Redfield Equation}, 
we find that they are mainly determined by the matrix elements 
$\bra{\rm in}\hat\sigma_\mu^a\ket{\rm in}$ and 
$\bra{\rm out}\hat\sigma_\mu^a\ket{\rm out}$, the others are 
exponentially suppressed. 

If we had $\bra{\rm in}\hat\sigma_\mu^a\ket{\rm in}=
\bra{\rm out}\hat\sigma_\mu^a\ket{\rm out}$, i.e., if the 
environment could not distinguish between the two states 
$\ket{\rm in}$ and $\ket{\rm out}$, then the 
effective Lindblad operators would be effective identity 
operators (in the subspace spanned by $\ket{\rm in}$ and 
$\ket{\rm out}$) and thus they would become trivial, i.e., 
the impact of the environment would vanish. 
However, for macroscopically different states $\ket{\rm in}$ 
and $\ket{\rm out}$, we generically have 
$\bra{\rm in}\hat\sigma_\mu^a\ket{\rm in}\neq
\bra{\rm out}\hat\sigma_\mu^a\ket{\rm out}$
for many matrix elements.
\new{This means that the environment can distinguish between the 
two states $\ket{\rm in}$ and $\ket{\rm out}$, unless 
there are cancellations between the different contributions.
These cancellations could be caused by precise 
relations between the environment operators.
This point is related to the concept of decoherence-free or 
symmetry-protected sub-spaces.} 
\new{In the absence of such cancellations,}
we again find  effective Lindblad operators 
in the form of projectors 
$\ket{\rm in}\bra{\rm in}$ and/or  $\ket{\rm out}\bra{\rm out}$.

Since the order of magnitude of the effective decay rate 
(or measuring frequency) $\Gamma$ 
as in Eqns.~\eqref{Gamma-Coarse} and \eqref{Gamma-Redfield}
is mainly determined by the properties of the environment 
and its coupling strength $g$ to the system, we \new{typically}  
find that it is small, but not exponentially suppressed.
In contrast, the transition rate of the system itself as 
determined by $\Delta E_{\rm min}=\exp\{-\ord(n)\}$ 
is exponentially small, such that we generically have 
$\Gamma\gg\Delta E_{\rm min}$ and thus we are again in 
the regime of the quantum Zeno effect. 

\section{Conclusions} 

The goal of this work is to study the impact of a quite general environment 
on the performance of adiabatic quantum algorithms, 
see also~\cite{childs2001a,tiersch2007a}.
To start with a specific example, we consider the adiabatic version of 
the Grover search algorithm (in a database with $N$ entries) 
proposed by Roland and Cerf~\cite{roland2002a}.
Apart from \new{a possible} environment induced shift of the system energies 
(which may cause problems for adaptive speed interpolations), 
the main impact of the environment is that it effectively 
monitors or measures the state of the system with a decoherence 
rate $\Gamma$. 
Since the minimum gap of the adiabatic quantum algorithm shrinks as 
$1/\sqrt{N}$ for large $N$ and thus the run-time $T$ grows as 
$\sqrt{N}$ whereas the properties of the environment 
(such as its temperature) and thus the decoherence rate $\Gamma$ 
do generically not display such a strong scaling behavior, 
we find $\Gamma T\gg1$ for large $N$.
As a result, the quantum Zeno effect inhibits or drastically slows down 
transitions from the initial state $\ket{s}$ to the desired final state 
$\ket{w}$. 

Note that already a single quasi-particle (photon, phonon or magnon etc.) 
in the environment whose interaction with only one of the qubits depends 
on the state of that qubit may induce an effective weak measurement. 
This problem is even more severe for the usual sequential quantum 
algorithms (without error correction), where one single measurement 
is sufficient to destroy 
the interference required for the algorithm to work. 
For adiabatic quantum algorithms, such a single measurement is not 
necessarily a problem, but frequently repeated measurements can 
inhibit or slow down the desired quantum evolution.  
Note that, even for comparable energies (i.e., in the vicinity of the 
minimum gap), environment-induced transitions between the states $\ket{s}$ 
and $\ket{w}$ (or other macroscopically different states) are quite 
unlikely since they imply changing the state of many qubits. 
In contrast, environment-induced effective (weak) measurements are much 
more likely since they do not require changing the state of many qubits 
-- they can be performed by acting on one qubit only. 

As one would already expect from the above arguments, analogous conclusions 
\new{should} 
apply to more general adiabatic quantum algorithms under certain assumptions 
(such as isolated Landau-Zener type avoided level crossings). 
As a bigger picture, the crucial point is to compare the effective
transition rate of the adiabatic quantum algorithm, analogous to 
\new{$\omega$} in the Introduction, with the measurement rate $\Gamma$ 
induced by the environment. 
Since the gap and thereby the effective transition rate 
\new{$\omega_{\rm sys}$}
generically decrease with increasing problem complexity 
or system size (number $n$ of qubits) while $\Gamma$ typically grows,
see also Eq.~\eqref{Gamma-Coarse}, one expects to enter the quantum Zeno 
regime \new{$\Gamma\gg\omega_{\rm sys}$} at some point.
Thus, while we did not prove that all adiabatic quantum algorithms suffer 
from this problem, our results strongly suggest to carefully consider  
the impact of an environment when trying to implement adiabatic quantum 
algorithms with many qubits (or other degrees of freedom).  

\section{Outlook} 

For the usual sequential quantum algorithms, the problem of decoherence 
as briefly sketched above has stimulated a substantial amount of work 
devoted to the development of error correcting codes such as the Shor 
code~\cite{shor1995a,steane1996a,laflamme1996a}. 
In the case of adiabatic quantum algorithms, it is \new{a bit} hard to see 
how such active error correcting codes could be applied directly, but 
one could envision other, more indirect, methods. 
One prominent example is the spin-echo technique where the overall 
phase of the system is frequently changed (e.g., reversed) without 
affecting the phase of the environment, such that the amplitudes 
of the interaction between system and environment  average out to 
zero if this phase-flip frequency is large enough, i.e., 
larger than effective environment frequency scales 
(such as the environment temperature).
Of course, by applying this idea, one must be careful to ensure that 
the repeated phase change of the system does not affect the adiabaticity 
of the internal system dynamics. 
Another option would be to encode the two states $\ket{s}$ and $\ket{w}$
\new{or $\ket{\rm in}$ and $\ket{\rm out}$}
within a decoherence-free \new{or symmetry-protected}
subspace such that the environment cannot 
distinguish between them (see also Sec.~\ref{Generalization}). 
\new{This point is related to the discussion in 
Sec.~\ref{Generalized Landau-Zener Transition}.
Even if the single-qubit expectation values are different 
$\bra{\rm in}\hat\sigma_\mu^a\ket{\rm in}\neq
\bra{\rm out}\hat\sigma_\mu^a\ket{\rm out}$, these 
differences could cancel out after summing over all environment 
operators -- provided that they are not independent of each other. 
A related effect could be achieved with structured noise models.}
Alternatively, one might consider monitoring the environment and using 
feedback control similar to quantum erasure schemes~\cite{wiseman2010}.

Furthermore, one could consider replacing the Landau-Zener like 
(tunneling) transition between macroscopically different quantum states 
such as $\ket{s}$ and $\ket{w}$ (which is very similar to a first-order 
phase transition) by a more gradual change of the state vectors from 
$\ket{s}$ to $\ket{w}$, e.g., similarly to a second-order transition~\cite{schuetzhold2006b,schaller2010a}. 
This may even lead to an enlarged minimum gap 
and thus a reduced run-time $T$.
In addition, such a gradual change should also alleviate 
the quantum Zeno mechanism sketched above. 
\new{In terms of the intuitive picture based on a complex potential 
landscape already discussed above, such a second-order phase 
transition would not correspond to the competition between two 
separated local minima (representing $\ket{s}$ and $\ket{w}$), 
but rather the smooth deformation of the global minimum into a 
pair of 
minima or a Mexican-hat shape or something similar.}

As a related point, the quantum Zeno effect considered here 
(and actually in most other scenarios) is applied to the 
(tunneling) transition between the two isolated states 
$\ket{s}$ and $\ket{w}$.  
If instead a continuum of states is involved, the character of the 
quantum Zeno effect changes and the required measurement frequency 
is typically much higher~\cite{ahmadiniaz2022a}.
In this case, the inverse decoherence rate $1/\Gamma$ should not be 
compared to the run-time $T$ but to the spreading time in the continuum,
which can be much shorter. 

\new{In order to place our work into context, let us also compare it to 
some alternative approaches. 
Here, we considered the competition between adiabatic quantum computation
and the quantum Zeno effect caused by the interaction with some environment.
Instead of such a competition, other works discussed how the quantum 
Zeno effect could be exploited in order to steer or engineer 
the quantum state or to actually perform the computation~\cite{berwald2024a,berwald2025a}. 
As discussed above, this could be achieved via active measurements or
passively by suitably coupling the system to some environment. 
As already mentioned above, these ideas are then also related to the 
widely discussed topics of 
quantum reservoir computing or dissipative quantum computing,
see, e.g., Refs.~\cite{verstraete2009b,kliesch2011a,sinayskiy2012a,fujii2021a,govia2021a}.}  

Apart from solving classical problems such as searching a database or 
factoring numbers, qubit architectures can also be used as quantum 
simulators for physical systems~\cite{biamonte2011a}.
For example, false vacuum decay has recently been simulated using 
several thousands of qubits~\cite{vodeb2025a}.
There are several similarities to the case considered here
(such as the analogy to first-order transitions), 
but also some important differences.
For example, false vacuum decay does typically not occur as a global 
transition between two states such as $\ket{s}$ and $\ket{w}$.  
Instead, it generally starts with a local nucleation seed 
(e.g., a few qubits are in the ``correct'' vacuum state while all 
the other are still in the false vacuum state) which then spreads out. 
Still, it would be interesting to study the impact of the environment,
e.g., regarding the quantum Zeno effect. 

In this context, quantum annealers or quantum annealing have also been 
discussed. 
Note that the term ``quantum annealing'' is not uniquely defined and can 
refer to different scenarios, ranging from quantum inspired classical 
algorithms to quantum algorithms which are also based on the adiabatic 
passage in a Landau-Zener type avoided level crossing. 
For the latter, the quantum Zeno effect induced by the environment 
should also be taken into account, as explained in the previous section.  

\ack{
Funded by the Deutsche Forschungsgemeinschaft 
(DFG, German Research Foundation) through 
the Collaborative Research Center
SFB~1242 ``Nonequilibrium dynamics of condensed matter
in the time domain'' (Project-ID 278162697).
R.S.~has benefited from the activities of COST Action CA23115: 
Relativistic Quantum Information, funded  by COST 
(European Cooperation in Science and Technology).}


\bibliographystyle{unsrt}
\bibliography{references}


\appendix 

\section{Freezing via Lindblad dissipation}\label{SEC:lindblad_slow}

We can write Eq.~\eqref{EQ:masterspec} also as $\frac{d}{dt} \expval{\f{\hat \sigma}} = M \expval{\f{\hat \sigma}}$ with matrix
\begin{align}\label{EQ:blochmat1}
M = \left(\begin{array}{ccc}
-2 \Gamma & 0 & 2 \omega\\
0 & -2 \Gamma & 0\\
-2 \omega & 0 & 0
\end{array}\right)\,,
\end{align}
and the eigenvalues of this matrix
$\lambda_0 = -2\Gamma$ and $\lambda_\pm = -\Gamma\pm\sqrt{\Gamma^2-4\omega^2}$ then support the discussion in the main text in the respective limits.

When we consider two hermitian Lindblad operators $\hat L_1 = \sqrt{\Gamma_1} \ket{\uparrow}\bra{\uparrow}$ or $\hat L_2 = \sqrt{\Gamma_2} \ket{\downarrow}\bra{\downarrow}$ instead, this changes into 
\begin{align}\label{EQ:blochmat2}
M = \left(\begin{array}{ccc}
-(\Gamma_1+\Gamma_2)/2 & 0 & 2 \omega\\
0 & -(\Gamma_1+\Gamma_2)/2 & 0\\
-2 \omega & 0 & 0
\end{array}\right)\,,
\end{align}
with analogous conclusions on the eigenvalues
$\lambda_0 = -(\Gamma_1+\Gamma_2)/2$ and $\lambda_\pm = (-\Gamma_1-\Gamma_2 \pm \sqrt{(\Gamma_1+\Gamma_2)^2-64 \omega^2})/4$.

Comparing~\eqref{EQ:blochmat1} and~\eqref{EQ:blochmat2} we see that a $\sigma^z$ Lindblad operator is more effective in dampening than e.g. a $\ket{\uparrow}\bra{\uparrow}$ Lindblad operator as it acts on both states.

In contrast, a non-hermitian Lindblad operator cannot be used to stabilize a state.
Considering for example the relaxation operator $\hat L=\sqrt{\sigma}\ket{\downarrow}\bra{\uparrow}$, we arrive at 
\begin{align}
M=\left(\begin{array}{ccc}
-\sigma/2 & 0 & 2 \omega\\
0 & -\sigma/2 & 0\\
-2\omega & 0 & -\sigma
\end{array}\right)\,,
\end{align}
which has eigenvalues $\lambda_0=-\sigma/2$ and $\lambda_\pm = (-3\sigma\pm \sqrt{\sigma^2-64\omega^2})/4$ which all maintain negative real parts in all regimes.

More generally, hermitian Lindblad operators can be associated with the Zeno effect.
In the eigen-basis of the measured observable, the net effect of repeated measurements is the erasure of coherences, which always increases the von-Neumann entropy of the measured system.
When we write~\eqref{EQ:master} with hermitian Lindblad operators $\hat L=\hat L^\dagger$ (multiple Lindblad operators work analogously) one can see
that such dephasing dissipators also never decrease the von-Neumann entropy of the system:
\begin{align}
\frac{d}{dt} S &= -\trace{\frac{d\hat\rho}{dt}\ln\hat\rho} = -\trace{(\hat L \hat\rho \hat L - \hat L^2 \hat \rho)\ln\hat\rho}\nn
&= \sum_{ab} L_{ab} L_{ba} \rho_a \ln \left(\frac{\rho_a}{\rho_b} \frac{L_{ab} L_{ba}}{L_{ab}L_{ba}}\right) \ge 0\,,
\end{align}
where in the first equality sign we have used that the term with the time derivative acting on the $\ln\hat\rho$ term does not contribute under the trace, 
in the second equality we exploited the invariance of the trace under cyclic permutations, 
and in the third equality we have evaluated the trace in the eigen-basis of the density matrix $\hat\rho=\sum_a \rho_a \ket{a}\bra{a}$ with $L_{ab} = \bra{a} \hat L \ket{b}$ and inserted the 
identity in the logarithm,
to eventually use the logarithmic sum inequality in the last step.

When additionally the Lindblad operators do not commute with the Hamiltonian, e.g. 
$\hat L=\sqrt{\Gamma} \hat O$ with $[\hat H, \hat O] \neq 0$, we can see an analogous slowing down of relaxation by dissipation:
Representing the density matrix $\hat \rho = \sum_{\ell m} \rho_{\ell m} \ket{\ell}\bra{m}$ now in the eigen-basis of the Lindblad operator defined via $\hat O \ket{\ell} = \lambda_\ell \ket{\ell}$, its matrix elements obey
\begin{align}
\dot\rho_{\ell m} &= -\ii \sum_n \rho_{n m} \bra{\ell}\hat H\ket{n}+\ii\sum_n \rho_{\ell n} \bra{n}\hat H\ket{m}
\new{- \frac{\Gamma}{2} (\lambda_\ell-\lambda_m)^2 \rho_{\ell m}}\,.
\end{align}
This implies that the diagonal elements just follow the evolution of the off-diagonals
$\dot\rho_{\ell\ell} = -\ii \sum_{n\neq \ell} \bra{\ell} \hat H \ket{n} \rho_{n\ell} + \ii \sum_{n\neq \ell} \bra{n} \hat H \ket{\ell} \rho_{\ell n}$, whereas the off-diagonal matrix elements will for large $\Gamma$ rapidly approach a flow equilibrium value (denoted by overbars) that is given by the diagonals
\begin{align}
\frac{\Gamma}{2} (\lambda_\ell-\lambda_n)^2 \bar\rho_{\ell n} 
\approx \ii \bra{\ell} \hat H \ket{n} (\rho_{\ell\ell}-\rho_{nn})\,.
\end{align}
By eliminating the off-diagonal matrix elements in the equation for the diagonals, we obtain a rate equation
\begin{align}
\dot \rho_{\ell\ell} \approx \sum_{n\neq \ell} \frac{4 \abs{\bra{\ell} \hat H \ket{n}}^2}{\Gamma (\lambda_\ell-\lambda_n)^2} \rho_{nn} - \left[\sum_{n\neq \ell} \frac{4 \abs{\bra{\ell} \hat H \ket{n}}^2}{\Gamma (\lambda_\ell-\lambda_n)^2}\right] \rho_{\ell\ell}\,,
\end{align}
which evolves slowly -- with rates of order $\ord\{\hat H^2/\Gamma\}$ -- towards a complete statistical mixture.
The stronger the dissipation in comparison to the system Hamiltonian, the slower is this secondary evolution, and as we have expressed the problem in the eigen-basis of $\hat L$, its eigen-states
are precisely the ones that are stabilized.

While Lindblad dynamics is typically observed in weak-coupling scenarios, we stress that the Zeno inhibition may be observed when $H_S$ is small in comparison to $\Gamma$. 
Furthermore, a frozen evolution due to strong dissipation can also be observed beyond Lindblad master equation scenarios as we show for a particular example in App.~\ref{SEC:general_slow}

\section{Freezing via strong dissipation}\label{SEC:general_slow}

To observe the Zeno effect beyond a weak-coupling scenario, we may consider a Caldeira-Leggett~\cite{caldeira1983a} type model of a frequency-driven oscillator
with position $\hat x$ and momentum $\hat p$ that is coupled to a reservoir of oscillators
\begin{align}
H(t) &= \frac{\hat p^2}{2m} + \frac{1}{2} m \omega^2(t) \hat x^2
\new{+ \sum_k \omega_k \left(\hat b_k^\dagger + \frac{h_k}{\omega_k}\hat x\right)\left(\hat b_k + \frac{h_k^*}{\omega_k} \hat x\right)}\,.
\end{align}
The operators then obey the Heisenberg equations of motion (we omit the time-dependence of the frequency below for brevity)
\begin{align}
\frac{d}{dt} \hat x(t) &= \frac{\hat p(t)}{m}\,,\nn
\frac{d}{dt} \hat p(t) &= - \left(m \omega^2 + 2\sum_k \frac{\abs{h_k}^2}{\omega_k}\right)\hat x(t)
\new{- \sum_k \left(h_k \hat b_k(t) + h_k^* \hat b_k^\dagger(t)\right)}\,,\nn
\frac{d}{dt} \hat b_k(t) &= -\ii \omega_k \hat b_k(t) -\ii h_k^* \hat x(t)\,.
\end{align}
In these, we may eliminate the reservoir modes and arrive at coupled equations for $\hat x(t)$ and $\hat p(t)$ only.
Introducing the spectral function of the reservoir $\Gamma(\omega) = 2\pi\sum_k \abs{h_k}^2\delta(\omega-\omega_k)$ and taking expectation values with the assumption of an initial product state with a reservoir state with $\expval{\hat b_k} = 0$ (e.g. a thermal one), we further obtain an integro-differential (non-Markovian) equation for position and momentum
\begin{align}
\frac{d}{dt} \expval{\hat p}_t &= - \left(m \omega^2 + \frac{1}{\pi} \int_0^\infty \frac{\Gamma(\omega)}{\omega} d\omega\right)\expval{\hat x}_t
\new{+ \int_0^t dt' \frac{1}{\pi} \int_0^\infty d\omega \Gamma(\omega) \sin(\omega(t-t')) \expval{\hat x}_{t'}}\,,\nn
\frac{d}{dt} \expval{\hat x}_t &= \frac{\expval{\hat p}_t}{m}\,.
\end{align}
Now, particularly for the spectral function
\begin{align}
\Gamma(\omega) = \Gamma_0 \frac{\omega\omega_c}{\omega^2+\omega_c^2}\,,
\end{align}
we observe that the memory kernel has a simple exponential dependence
\begin{align}
\frac{d}{dt} \expval{\hat p}_t &= - \left(m \omega^2 + \frac{\Gamma_0}{2}\right)\expval{\hat x}_t 
+ \int\limits_0^t dt' \frac{\Gamma_0\omega_c}{2} e^{-\omega_c (t-t')}\expval{\hat x}_{t'}\,,\nn
\frac{d}{dt} \expval{\hat x}_t &= \frac{\expval{\hat p}_t}{m}\,.
\end{align}
This form has the tremendous advantage that also for finite $\omega_c$ we may alternatively solve the time-local enlarged system
\begin{align}
\frac{d}{dt} \expval{\hat p}_t &= -\left(m \omega^2 + \frac{\Gamma_0}{2}\right) \expval{\hat x}_t + B_t\,,\nn
\frac{d}{dt} \expval{\hat x}_t &= \expval{\hat p}_t/m\,,\nn
\frac{d}{dt} B_t &= \frac{\Gamma_0 \omega_c}{2} \expval{\hat x}_t - \omega_c B_t
\end{align}
with the initial condition $B_0=0$.

We can introduce the dimensionless variables $\tilde x = \sqrt{2m \omega} \expval{\hat x}_t$, $\tilde p = \sqrt{2/(m \omega)} \expval{\hat p}_t$, $\tilde B = \sqrt{2/(m \omega^3)} B_t$ and the dimensionless time $\tau=\omega t$, such that the vector $\f{r} = (\tilde x, \tilde p, \tilde B)$ obeys $\left(1+\frac{\tau\dot\omega}{\omega^2}\right)\frac{d}{d\tau} \f{r}(\tau) = M \f{r}(\tau)$ with the coefficient matrix
\begin{align}
M = \left(\begin{array}{ccc}
0 & 1 & 0\\
-(1+\alpha) & 0 & 1\\
\alpha \beta & 0 & -\beta
\end{array}\right)\,,
\end{align}
where $\alpha = \frac{\Gamma_0}{2m \omega^2} \ge 0$ denotes a dimensionless damping strength and $\beta = \frac{\omega_c}{\omega} \ge 0$ a dimensionless reservoir memory scale.
Thus, the time-local evolution is governed by the $\tau$-dependent eigenvalues of $M$, and the strong-coupling regime can be reached by small $\omega$.
The eigenvalues of $M$ all have a non-positive real part and can be expressed in terms of radicals.
Furthermore, beyond a certain damping strength $\alpha>8$ and for $\frac{\sqrt{\alpha^2+20\alpha-8-\sqrt{\alpha}(\alpha-8)^{3/2}}}{2^{3/2}} < \beta < \frac{\sqrt{\alpha^2+20\alpha-8+\sqrt{\alpha}(\alpha-8)^{3/2}}}{2^{3/2}}$ all eigenvalues are just real negative. 
When $\beta = 3 \sqrt{\frac{\alpha-2}{2}}$ and $\alpha\ge 8$, the eigenvalues read
\begin{align}
\lambda_1 &= \frac{\sqrt{\alpha-8}-\sqrt{\alpha-2}}{\sqrt{2}}\,,\qquad
\lambda_2 = -\frac{\sqrt{\alpha-2}}{\sqrt{2}}\,,\nn
\lambda_3 &= -\frac{\sqrt{\alpha-8}+\sqrt{\alpha-2}}{\sqrt{2}}\,,
\end{align}
from which we see that the first tends to zero $\lim_{\alpha\to\infty} \lambda_1 = -3/\sqrt{2\alpha}$ (Zeno stabilization). 
The corresponding right eigenvector is given by
\begin{align}
v_1 = \left(\begin{array}{c}
1\\
\lambda_1\\
\alpha+1+\lambda_1^2
\end{array}\right)\frac{1}{\sqrt{1+\lambda_1^2+(\alpha+1+\lambda_1^2)^2}}\,,
\end{align}
which shows that states with a negligible momentum component will be stabilized for large $\alpha$ (small $\abs{\lambda_1}$).
Classically, this Zeno inhibition just corresponds to the overdamped limit, where momentum first settles rapidly to a flow equilibrium value and then position relaxes extremely slowly.


\section{Adiabatic master equation}

In the adiabatic master equation of Lindblad form, 
one assumes strictly adiabatic time evolution~\eqref{EQ:adiabatic} and performs the Born- and Markov approximations as usual.
This results in a time-dependent Redfield equation that is not of Lindblad form and thereby only preserves trace and hermiticity
but not positivity of the system density matrix.
To obtain a Lindblad form in general, a secular approximation is already required for undriven systems.
Analogously, for adiabatically driven systems, a generalized secular approximation can be performed that eventually leads to a Lindblad master equation, 
see Eqns.~(54) in Ref.~\cite{albash2012a}.
Technically, one could obtain the adiabatic master equation also from the time-independent problem (see e.g. Eq.~(32) from Ref.~\cite{landi2022a}) by inserting the time-dependencies of eigenvalues and eigen-states {\em a posteriori}.
Thus, the favorable properties of the Born-Markov-secular master equation also transfer to the adiabatic version and one obtains a decoupled evolution of populations and coherences for non-degenerate systems.
The populations then follow a Pauli-type rate equation as e.g. given by Eq.~(36) of Ref.~\cite{landi2022a}, but where the time-dependent rates are evaluated at the instantaneous energies.
This then also implies that a Zeno blockade cannot be observed in the adiabatic master equation as the ground state will always be favored when the temperature is not much larger than the minimum gap. 
Consistently, the Zeno blockades observed earlier by some of us~\cite{schaller2018a} have been obtained using master equations that do not decouple populations and coherences in the energy eigen-basis.

Thus, we object that the generalized secular approximation will fail when the driving protocol leads to small energy gaps (e.g. the shaded regions in Fig.~\ref{FIG:groverspectrum}) 
as is unfortunately the case for most adiabatic algorithms.
Consequently, the results from the associated adiabatic master equation (and resulting Pauli rate equation) should only be trusted in regimes where the gap remains large (white regions in Fig.~\ref{FIG:groverspectrum}).

\section{Singular coupling master equation}

When the gap is negligible in comparison to the system-reservoir coupling strength (deep within the shaded regions in Fig.~\ref{FIG:groverspectrum}), one can also derive a master equation by treating both $\hat H_{\rm sys}(t)$ and $\hat H_{\rm int}$ as perturbation.
This is then possible as in the relevant near-degenerate subspace the system Hamiltonian (up to a trivial shift) only opens a small gap and is thus small in comparison to the interaction.
This scenario is analogous to the singular coupling limit~\cite{palmer1977a}.
More explicitly, starting from $\hat H_{\rm int}=g \sum_\alpha \hat A_\alpha \otimes \hat B_\alpha$ (the system operators $\hat A_\alpha$ and reservoir operators $\hat B_\alpha$ can always be chosen individually hermitian), the interaction picture Hamiltonian becomes
\begin{align}
\hat H(t) &= \hat H_{\rm sys}(t) + g \sum_\alpha \hat A_\alpha \otimes \hat B_\alpha(t)\,,\nn
\hat B_\alpha(t) &= e^{+\ii \hat H_{\rm env} t} \hat B_\alpha e^{-\ii \hat H_{\rm env} t}\,,
\end{align}
and by iteratively solving the von-Neumann equation neglecting terms of $\ord\{H_{\rm int}^3\}$ and $\ord\{H_{\rm int} H_{\rm sys}(t)\}$ and performing the Born and Markov approximations as before, we arrive at (this equation is the same in interaction and Schr\"odinger pictures)
\begin{align}
\frac{d\hat\rho_{\rm sys}}{dt} &= -\ii \left[\hat H_{\rm sys}(t), \hat \rho_{\rm sys}\right]\nn
&\qquad- g^2 \sum_{\alpha\beta} \int_0^\infty d\tau C_{\alpha\beta}(+\tau) \left(\hat A_\alpha, \hat A_\beta \hat \rho_{\rm sys}\right]
\new{- g^2 \sum_{\alpha\beta} \int_0^\infty d\tau C_{\beta\alpha}(-\tau) \left[\hat \rho_{\rm sys} \hat A_\beta, \hat A_\alpha\right]}\nn
&= -\ii \left[\hat H_{\rm sys}(t) + \frac{g^2}{2\ii} \sum_{\alpha\beta} \sigma_{\alpha\beta} \hat A_\alpha \hat A_\beta, \hat \rho_{\rm sys}\right]
\new{+g^2\sum_{\alpha\beta} \gamma_{\alpha\beta} \left[\hat A_\beta \hat \rho_{\rm sys} \hat A_\alpha - \frac{1}{2}\left\{\hat A_\alpha \hat A_\beta, \hat \rho_{\rm sys}\right\}\right]}\,,
\end{align}
where $\sigma_{\alpha\beta}=\int_{-\infty}^\infty C_{\alpha\beta}(\tau){\rm sgn}(\tau) d\tau$ and $\gamma_{\alpha\beta}=\int_{-\infty}^\infty C_{\alpha\beta}(\tau) d\tau$.
As the matrix $\gamma_{\alpha\beta}$ is positive semi-definite due to Bochners theorem, this is a Lindblad form.
We can also see that it has the completely mixed state as a stationary solution.
Furthermore, the above equation is identical to the singular-coupling master equation with time-dependencies inserted {\em a posteriori}, compare e.g. Eq.~(48) of Ref.~\cite{landi2022a}.

Thus, we see that in the regime of small energy gaps in the relevant subspace, the hermitian coupling operators directly transfer to hermitian Lindblad operators that can be interpreted as a measurement performed by the environment. 
For undriven systems, combinations of small and large energy gaps have been treated before~\cite{schultz2009a,trushechkin2021a}, but for driven systems this has to be generalized.

\end{document}